\documentclass[onecolumn]{aa} 

\usepackage{graphicx}
\usepackage{txfonts}
\usepackage{multirow}
\usepackage{natbib,twoopt}
\usepackage[breaklinks=true]{hyperref} 
\bibpunct{(}{)}{;}{a}{}{,}             
\makeatletter
\newcommandtwoopt{\citeads}[3][][]{\href{http://adsabs.harvard.edu/abs/#3}%
	{\def\hyper@linkstart##1##2{}%
		\let\hyper@linkend\@empty\citealp[#1][#2]{#3}}}
\newcommandtwoopt{\citepads}[3][][]{\href{http://adsabs.harvard.edu/abs/#3}%
	{\def\hyper@linkstart##1##2{}%
		\let\hyper@linkend\@empty\citep[#1][#2]{#3}}}
\newcommandtwoopt{\citetads}[3][][]{\href{http://adsabs.harvard.edu/abs/#3}%
	{\def\hyper@linkstart##1##2{}%
		\let\hyper@linkend\@empty\citet[#1][#2]{#3}}}
\newcommandtwoopt{\citeyearads}[3][][]%
{\href{http://adsabs.harvard.edu/abs/#3}
	{\def\hyper@linkstart##1##2{}%
		\let\hyper@linkend\@empty\citeyear[#1][#2]{#3}}}
\makeatother

\begin{document}

   \title{Photometry, spectroscopy, and polarimetry of distant comet C/2014 A4 (SONEAR)}

   \author{Oleksandra Ivanova\inst{1,2,3}\fnmsep\thanks{Corresponding author, \email{oivanova@ta3.sk}},
		Igor Luk’yanyk\inst{3},
		Ludmilla Kolokolova\inst{4},
		Himadri Sekhar Das\inst{5},
		Marek Hus\'arik\inst{1},
		Vera Rosenbush\inst{2,3},
		Viktor Afanasiev\inst{6},
		J\'an Svore\u n\inst{1},
		Nikolai Kiselev\inst{2,7}
    \and
        Vadim Krushinsky\inst{8}
}
   \institute{
  	Astronomical Institute of the Slovak Academy of Sciences, SK-05960 Tatransk\'a Lomnica, Slovak Republic
  	\and
  	Main Astronomical Observatory of the National Academy of Sciences, 27 Akademika Zabolotnoho St., 03143 Kyiv, Ukraine
  	\and
  	Astronomical Observatory, Taras Shevchenko National University of Kyiv, 3 Observatorna St., 04053 Kyiv, Ukraine
  	\and
  	University of Maryland, College Park, Maryland, USA 
  	\and
  	Department of Physics, Assam University, Silchar 788011, Assam, India
  	\and
  	Special Astrophysical Observatory, Russian Academy of Sciences, Nizhnii Arkhyz, 369167 Russia
  	\and
  	Crimean Astrophysical Observatory, Nauchnij, Crimea
  	\and
  	Ural Federal University, Ekaterinburg, 620083 Russia
  }
  
  \date{Received January 21, 2018}
  
  
  \abstract
  {The study of distant comets, which are active at large heliocentric distances, is important for better understanding of their physical properties and mechanisms of long-lasting activity. }
  {We analyze the dust environment of the distant comet C/2014 A4 (SONEAR), with a perihelion distance near 4.1~au, using comprehensive observations obtained by different methods.}
  {We present an analysis of spectroscopy, photometry, and polarimetry of comet C/2014 A4 (SONEAR), which were performed on November 5~--~7, 2015, when its heliocentric distance was 4.2~au and phase angle was 4.7$^\circ$. Long-slit spectra and photometric and linear polarimetric images were obtained using the focal reducer SCORPIO-2 attached to the prime focus of the 6-m telescope BTA (SAO RAS, Russia). We simulated the behavior of color and polarization in the coma presenting the cometary dust as a set of polydisperse polyshapes rough spheroids.}
  {No emissions were detected in the 3800~--~7200~$\AA$ wavelength range. The continuum showed a reddening effect with the normalized gradient of reflectivity 21.6$\pm$0.2\% per 1000~$\AA$ within the 4650~--~6200~$\AA$ wavelength region. The fan-like structure in the sunward hemisphere was detected. The radial profiles of surface brightness differ for r-sdss and g-sdss filters, indicating predominance of submicron and micron-sized particles in cometary coma. The dust color (g--r) varies from 0.75$ \pm $0.05$^m$ to 0.45$ \pm $0.06$^m$ along the tail. For aperture radius near 20~000~km, the dust productions in various filters were estimated as $Af\rho $~=~680$\pm$18~cm (r-sdss) and 887$ \pm $16~cm (g-sdss). The polarization map showed spatial variations of polarization over the coma from about --3\% near the nucleus to --8\% at cometocentric distance about 150~000~km. Our simulations show that the dust particles were dominated (or covered) by ice and tholin-like organics. Spatial changes in the color and polarization can be explained by particle fragmentation. }
  {}
  
  
  \titlerunning{Distant comet C/2014 A4 (SONEAR)}
  \authorrunning{Ivanova et al.}
  
  \maketitle
%

\section{Introduction}
Small ice-rich bodies from the outer region of the Solar System are recognized as remnants of the long-time evolution of the protoplanetary nebula. Currently, three reservoirs of the comets are mainly identified: the trans-Neptunian region, the Oort Cloud, and Asteroid belt (i.e., main-belt comets and active asteroids) \citepads{2015SSRv..197..191D}. Dynamical studies revealed that (i) the Oort Cloud is a primary source of "nearly isotropic comets", including long-period (LP) comets and Halley-type comets, in turn (ii) the flattened Kuiper Belt (ecliptic comets and Encke-type comets) and (iii) the Scattered disk --- for the Jupiter-family comets (JFCs) and Centaurs \citepads{1990Natur.344..825W,1997NYASA.822...67W,2010Sci...329..187L}. But from the dynamical evolutionary scenarios, it is not clearly understood if short- and long-period comets were formed at the distinct places or in the slightly overlapped regions of the primordial planetesimal disk before they were scattered to the outer region of the Solar System \citepads{2004ASPC..323..371D,2015SSRv..197..191D,2008Icar..196..258L,2017ApJ...845...27N}.

Remote investigations of primitive matter in the Solar System have traditionally been carried out through observations of long-period comets, which are less affected by solar radiation than their short-period counterparts orbiting much closer to the Sun. A dynamically new comet is usually defined as a comet with the semi-major axis of orbit $a > 10 000$ au, or orbital period $P > 1$ million yr \citepads{2013come.book.....E,2014A&A...565A..69M}. They reside very far from the Sun and spend the majority of their lifetime even outside the heliosphere. Therefore, these comets are exposed to marginal solar radiation and are only very little modified by the solar wind. In spite of this, dynamically new comets are characterized by a higher level of activity on average \citepads{2009Icar..201..719M,2015Icar..258...28I,2016AJ....152..220S}.\\

Currently for distant comets, molecular bands in spectra above the underlying continuum were detected in two comets – 29P/Schwassmann-Wachmann 1 at heliocentric distance about 5.9~au and C/2002 VQ94 (LINEAR) at 7.3~au \citepads[][and references therein]{2006A&A...459..977K,2008Icar..198..465K}. Tails were not observed in these comets, and their morphologies are characterized by asymmetric comae with jet features \citepads{2004ApJS..154..463S,2006A&A...459..977K,2008Icar..198..465K,2008A&A...485..599T,2016P&SS..121...10I,2019Icar..319...58P}. Other distant comets did not show gas emissions above the reflected solar continuum in the optical spectra, although some of these comets demonstrated long dust tails \citepads{1962PASP...74..537R,1965AJ.....70..451B,2003A&A...410.1029K,2009Icar..201..719M,2015P&SS..118..199I,2015Icar..258...28I}. 

The compositional taxonomy of the small icy bodies allows establishing a possible correlation between physical properties of the distant dwellers and their orbital characteristics shedding light on the primordial physical conditions at the places of their formation \citepads{2015SSRv..197..297M}. Studies of comets in their active phase can be especially informative. It is especially interesting to trace the activity evolution of the dynamically new comets moving into the inner Solar System for the first time and compare it with the activity of returning comets, which have had passages through the inner Solar System and as a result have their outermost layers differentiated and depleted of volatiles. 

 The main mechanisms that have been proposed to explain the activity of comets at large heliocentric distances are: the phase transition  between amorphous and crystalline water ice \citepads{1992ApJ...388..196P,2002EM&P...90..217C}, the annealing of amorphous water ice \citepads{2009Icar..201..719M}, and the sublimation of more volatile components like CO and/or CO$_2$. For comets, which are active at heliocentric distances more than 4~au and where the temperature of nucleus surface is lower than the H$_2$O sublimation temperature, the activity is likely supported by the sublimation of more volatile CO$_2$ that was tentatively confirmed by the AKARI observations \citepads{2012ApJ...752...15O} or CO \citepads{2018arXiv181107180J}. However, as shown by \citetads{2011Icar..211..559I}, the upper layers  of the cometary nucleus must be peeled off continually in order to satisfy the observational results for a long-term high activity of the comets and the most popular idea of crystallization of amorphous water ice cannot be responsible for the observed long-term cometary activity at large heliocentric distances.

The volatile molecule abundances do not show any correlation with dynamical families, leading to a conclusion that the both groups of comets were formed in largely overlapping regions \citepads{2012ApJ...758...29A}. Contrary to the color diversity in the Kuiper Belt population, no difference was found between the mean colors of the short- and long-period comets confirming the absence of significant composition differences between these two groups \citepads{2015AJ....150..201J}. Although no difference was found between the mean colors of dynamically new and returning long-period comets \cite{2012Icar..218..571S}, it is very important to continue studying comets from different dynamic groups.

In this work, we investigate one of the new LP comets C/2014 A4 (SONEAR) (hereafter C/2014 A4). In CBET \#3783, issued on January 16, 2014, it was announced a discovery of an apparently asteroidal object ($\sim$18.1$^m$) by Cristovao Jacques, Eduardo Pimentel, and Joao Ribeiro de Barros on CCD images obtained on January, 2014, 12.0 UT with a 0.30-m f/3 reflector of the Southern Observatory for Near Earth Research (SONEAR) at Oliveira, Brazil. The object showed a cometary appearance at perihelion distance 4.18~au  \cite{Williams}.

We present an analysis of optical observations of comet C/2014 A4 at a post-perihelion heliocentric distance 4.2~au. The paper is organized as follows: observations and data reduction are presented in Section 2. In section 3 we analyze all the data, including polarimetry, Section 4 represents modeling of polarization and color, and Section 5 summarizes our conclusions.
      
\section{Observations and data reduction}

The observations of C/2014 A4 were made with the 6-m telescope BTA of the Special Astronomical Observatory (Russia) on November 5, 2015, when heliocentric and geocentric distances of the comet were 4.2 and 3.3~au, respectively, and phase angle was 4.7$^{\circ}$. The focal reducer SCORPIO-2 attached to the prime focus of the telescope was operated in the photometric, polarimetric, and long-slit spectroscopic modes \citepads{2011BaltA..20..363A}. A CCD chip E2V42-90 consisting of 4600$\times$2048 pixels was used as a detector. The size of one pixel is 13.5$\times$13.5~$\mu$m which corresponds to 0.18$^{\prime\prime}\times$0.18$^{\prime\prime}$ on the sky plane. 

The photometric data of C/2014 A4 were obtained through the g-sdss (the central wavelength $\lambda_0$ and FWHM are represented as $\lambda$4650/650~$\AA$) and r-sdss ($\lambda$6200/600~$\AA$) broadband filters. The seeing was stable around 1.4$^{\prime\prime}$. Performing photometric and polarimetric measurements, we applied a 2$\times$2 binning of the original frames. The dimension of the images was 1024$\times$1024~px and the scale was 0.36$^{\prime\prime}$/px. The full field of view of the CCD is 6.1$^{\prime} \times$ 6.1$^{\prime}$. Bias subtraction, flat field correction, and cleaning from cosmic ray tracks were made. The sky background for each individual frame was estimated from those parts of the frame that were not covered by the cometary coma and free of faint stars, using procedure of building histogram of counts in the image. Because measurements were differential, we combined the frames of the comet using only central contour of isophotes closest to the maximum of the comet brightness. 

Spectroscopic observations were made with a long-slit mask. The height of the slit was 6.1$^{\prime}$ and the width was 1$^{\prime\prime}$. The transparent grism  \href{https://www.sao.ru/hq/lsfvo/devices/scorpio-2/grisms_eng.html}{VPHG1200@540} was used as a disperser in the spectroscopic mode. The spectra covered the wavelength range of 3800~---~7200~$\AA$ and had a dispersion of 0.81~$\AA$/px. The spectral resolution of the spectra was defined by the width of the slit and was about 5.2~$\AA$. The spectroscopic images were binned along the spatial direction as 1$\times$2. We subtracted the bias, performed the flat-fielding, and removed the hits of cosmic rays. The wavelength calibration was made using the spectrum of He-Ne-Ar lamp. After that we performed the linearization and summation of the spectra obtained with the same slit position. To estimate a night sky spectrum, we used the peripheral regions of the slit where the comet light contribution was negligible. Then we transformed the peripheral night sky spectrum into the Fourier space and extrapolated it onto the comet position by using the polynomial representation at a given wavelength. The telescope was tracked on the comet to compensate its apparent motion during the exposure.

To provide the absolute photometric and spectral calibration we observed standard star G191-B2B \citepads{1990AJ.....99.1621O} and used the measurements of the spectral atmospheric transparency at the Special Astronomical Observatory provided by \citetads{1978AISAO..10...44K}. Standard reduction of the obtained spectroscopic data was made. To remove biases from the observed frames, to clean the frames from the cosmic events, and to correct their geometry we used specialized software packages in the IDL environment developed at the SAO RAS. 
\begin{table*}[t]
	\centering
	\caption[]{Log of the observations of comet C/2014 A4 (SONEAR).}
	\label{tab:1} 
	\begin{tabular}{lccccccccc}
		\hline
		\noalign{\smallskip}
		Date, UT & $r$ & $\Delta$ & $\alpha$ & $\phi$ & Filter/grism & $T_{exp}$ &  $N$ &  Mode & Telescope \\
		& (au)&   (au)   & (deg)    & (deg)    & &    (s)    &      &  & \\
		\noalign{\smallskip}
		\hline
		\noalign{\smallskip}
		2015 Nov. 5.84 & 4.21 & 3.28 & 4.7 & 85.2 & g-sdss & 30 & 5 & Ima & BTA SAO\\
		2015 Nov. 5.85 & 4.21 & 3.28 & 4.7 & 85.2 & r-sdss & 30 & 5 & Ima & 	–"–\\
		2015 Nov. 5.86 & 4.21 & 3.28 & 4.7 & 85.2 & Rc & 30 & 15 & ImaLP & –"–\\
		2015 Nov. 5.87 & 4.21 & 3.28 & 4.7 & 85.2 & VPHG1200@540 & 600 & 3 & Sp & –"–\\
		2015 Nov. 7.76 & 4.21 & 3.29 & 5.8 & 83.6 & V & 180 & 10 & Ima & MASTER-II\\
		2015 Nov. 7.78 & 4.21 & 3.29 & 5.8 & 83.6 & R & 180 & 10 & Ima & –"–\\
		\noalign{\smallskip}
		\hline
	\end{tabular} 	
\end{table*}

The polarimetric data were obtained with the broadband Rc ($\lambda$6420/790~$\AA$) filter of the Johnson-Cousins photometric system. For measurements of the degree of linear polarization of the comet, we used the dichroic polarization analyzer (POLAROID) \citepads{2012AstBu..67..438A}. The reduction of the obtained polarimetric images was performed using standard techniques of bias frame subtraction, flat-field correction, and preparation of images for processing. The corrected frames were stacked. We also observed the polarized and non-polarized standard stars from the lists of \citetads{1982ApJ...262..732H}, \citetads{1992AJ....104.1563S}, and \citetads{2000AJ....119..923H}  for estimation of the instrumental polarization. Typically the instrumental polarization was less than 0.1\%. A detailed description of the data reduction and the method of the calculation of polarization parameters with SCORPIO-2 can be found in \citetads{2012AstBu..67..438A},   \citetads{2014AstBu..69..211A},  \citetads{2009SoSyR..43..453I,2017Icar..284..167I}, and \citetads{2017MNRAS.469S.475R}. The errors in the averaged polarization degree varied from  0.12\% in the near-nucleus region to 1.1\% in the tail at projected distance $\rho$ from the coma photocenter $\sim$150~000~km.

Also, the photometric observations of comet C/2014 A4 were carried out at the 0.4-m MASTER-II-Ural telescope (Kourovka observatory of the Ural Federal University, Russia) on November 7, 2015. Comet was observed through Johnson’s broadband V ($\lambda$5510/880~$\AA$) and R ($\lambda$6580/1380~$\AA$) filters. The CCD Apogee ALTA U16M CCD was used as the detector with an image size of 4096$\times$4096~px and a scale of 1.85$^{\prime\prime}$/px.
 
The viewing geometry and log of observations of comet C/2014 A4 are presented in Table \ref{tab:1} that lists the mid-cycle time, the heliocentric ($ r $) and geocentric ($\Delta$) distances, the phase angle of the comet ($\alpha$), the position angle of the extended Sun-comet radius vector ($\phi$), the filter/grism, the total exposure time during the night ($T_{exp}$), the number of cycles of observations obtained during night ($ N $), the mode of the observation: Images (Ima), Images of Linear Polarization (ImaLP), or Spectra (SP), and telescopes.
\begin{figure*}[b!]
	\centering
	\includegraphics[width=0.8\linewidth]{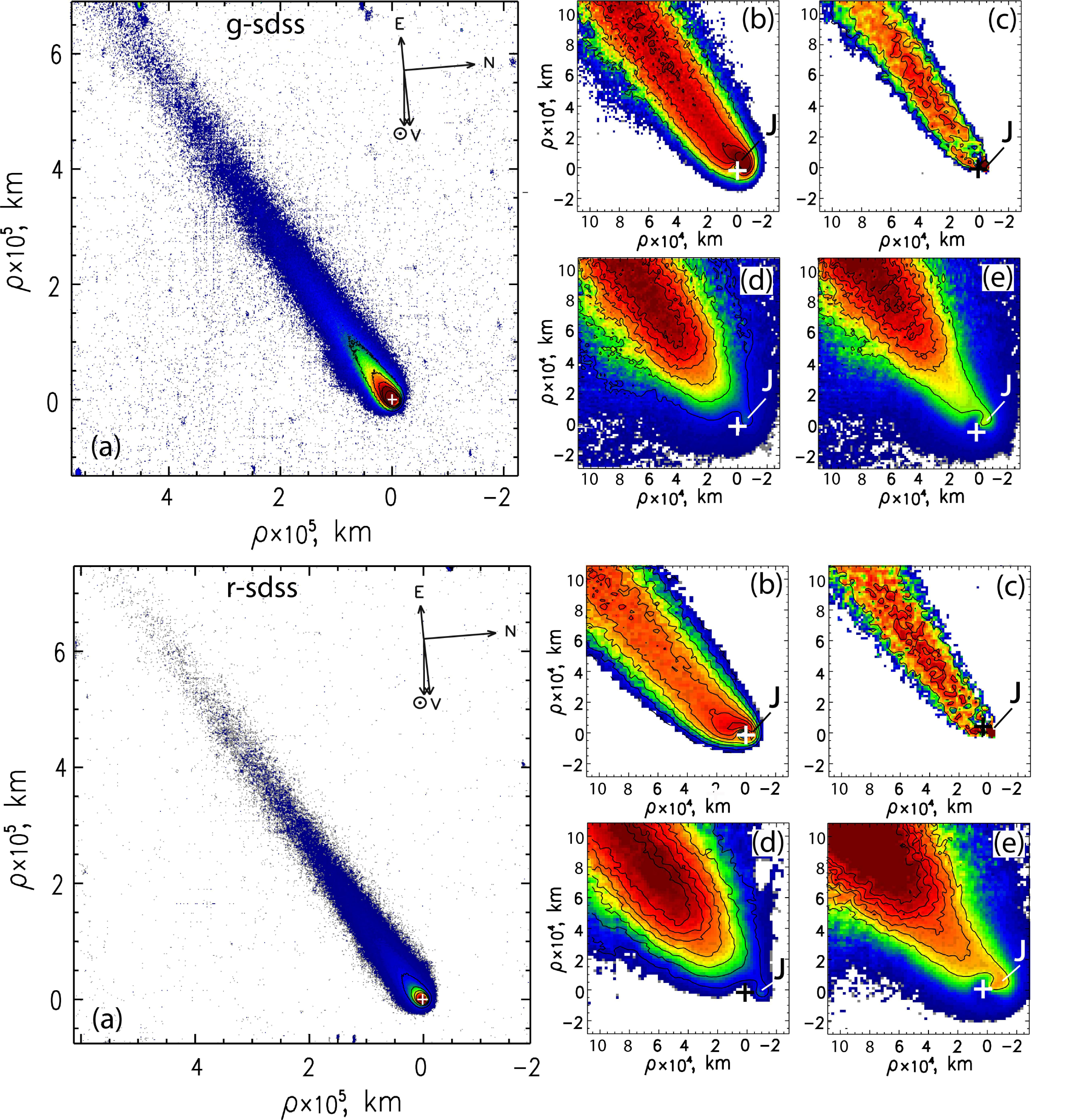}
	\caption{Images of comet C/2014 A4 (SONEAR) in the g-sdss filter (top panel) and r-sdss filter (bottom panel), obtained at the BTA/SCORPIO-2. (a) shows the direct images of the comet in relative intensity; (b) is the relative intensity image processed using the division by the 1/$\rho$ profile method \citepads{2014Icar..239..168S}; (c) is the image processed by a rotational gradient method \citepads{1984AJ.....89..571L}; (d) and (e) represent the relative intensity images to which a division by azimuthal average and renormalization methods were applied, respectively \citepads{2014Icar..239..168S}. The relative intensities of adjacent contours of isophotes differ by factor of $\sqrt{2}$. Color scale does not reflect the absolute brightness of the comet, and is used only to separate the areas of intensities that differ by $\sqrt{2}$. $\rho$ is a distance from cometary optocenter. J is a fan-like structure in the near-nucleus coma. The arrows point in the direction to the Sun ($\odot$), North (N), East (E), and velocity vector of the comet as projected onto the sky plane (V). }
	\label{Fig:01}
\end{figure*}

\section{Photometry: analysis of the observed data}

\subsection{Morphology}
For the morphology analysis, we used the images of comet C/2014 A4 obtained through the broadband g-sdss ($\lambda$4650/650~$\AA$) and r-sdss ($\lambda$6200/600~$\AA$) filters at the 6-m telescope SAO RAS. Fig. \ref{Fig:01} shows direct images of comet C/2014 A4 with relative isophots (a) and images treated with digital filters (b-e). To highlight low-contrast structures in the images, we have applied a combination of numerical techniques: a rotational gradient method \citepads{1984AJ.....89..571L}, Gauss blurring, division by azimuthal average, 1/$\rho$ profile method, and median filtering \citepads{2014Icar..239..168S}, as shown in Fig. \ref{Fig:01}. To exclude spurious features when interpreting the obtained images, each of the digital filters was applied with the same processing parameters for all individual exposures. This technique was used before to pick out structures in several comets with good results \citepads{2009SoSyR..43..453I,2016P&SS..121...10I,2017Icar..284..167I,2018Icar..313....1I,2017MNRAS.469S.475R,2019Icar..319...58P}. 
	
The comet displayed an extended coma with highly condensed material in the near-nucleus area and a long tail in the antisolar direction, typical for most comets active at large heliocentric distances \citepads{2003A&A...410.1029K,2008Icar..198..465K,2015P&SS..118..199I,2015Icar..258...28I,2016AJ....152..220S}. We can see in Fig. \ref{Fig:01} that there are no features in the tail, but there is a fan-like structure in the sunward hemisphere which is denoted as J. 
 
\begin{figure}[t]
	\centering
	\includegraphics[width=1.0\linewidth]{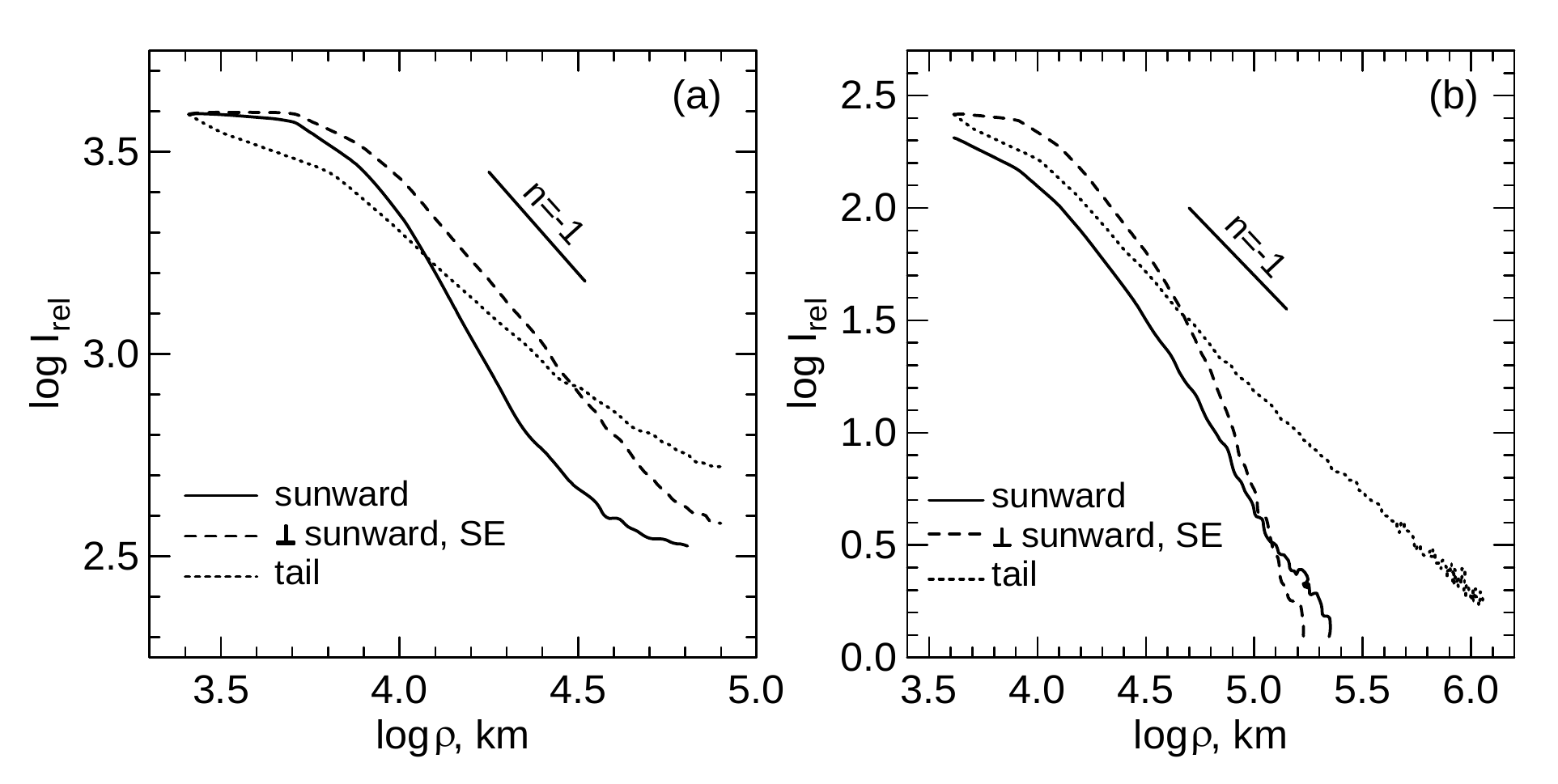}
	\caption{Observed surface brightness profiles of comet C/2014 A4 (SONEAR)  for flux calibrated images acquired in g-sdss (a) and r-sdss (b) bands. The individual curves are the scans taken through the photometric centre in different directions. }
	\label{Fig:02}
\end{figure}

\subsection{Radial surface brightness profiles}
For analysis of the radial surface brightness profile of the coma and tail, we used the g-sdss and r-sdss images of comet C/2014 A4. The individual curves in Fig. \ref{Fig:02} show variations of the average coma flux in the 3$\times$3 px size aperture with increasing distance from the nucleus. The cuts across the intensity map of the comet are made in different directions: along (solid line) and perpendicular (dash line) to the solar direction, and along the tail (dotted line), the position angle of which, measured counterclockwise from the direction to North through East, is 121$^\circ$. Fig. \ref{Fig:02} shows large differences between the profiles of tail and coma as well as between the profiles in the sunward and perpendicular to the solar direction. There is a gentle decline of the surface brightness in the tail, while a sharp drop of brightness is seen in the coma.

\begin{table}[b]
	\centering
	\caption[]{Radial slopes of profiles, in a log-log scale, measured in the g-sdss and r-sdss images of comet C/2014 A4 (SONEAR) at different ranges of coma distances.}
	\label{tab:2} 
	\begin{tabular}{cc|cc}
		\hline
		\multicolumn{2}{c}{g-sdss} & \multicolumn{2}{c}{r-sdss} \\
		\noalign{\smallskip}
		Distance range  & Slope & Distance range & Slope \\
		(km) &  & (km) &   \\
		\noalign{\smallskip}
		\hline
		\noalign{\smallskip}
		\multicolumn{4}{c}{Tail direction } \\
		2570–6457   & –0.36 $\pm$ 0.03 & 	5012–10233 & –0.46 $\pm$ 0.10 \\
		6457–28840  & –0.79 $\pm$ 0.01 &   10233–97724 & –1.05 $\pm$ 0.02 \\
		28840–67608 & –0.54 $\pm$ 0.01 &               &                  \\
		\noalign{\smallskip}
		\multicolumn{4}{c}{Sunward direction } \\
		\noalign{\smallskip}
		7586–20417  & –1.29 $\pm$ 0.04 & 	8318–16596 & –0.76 $\pm$ 0.10 \\
		20417–38019 & –0.95 $\pm$ 0.03 & 16596–69183   & –1.47 $\pm$ 0.02 \\
		\noalign{\smallskip}
		\multicolumn{4}{c}{Perpendicular to the sunward direction, SE } \\
		\noalign{\smallskip}
		7762–56234	& –1.04 $\pm$ 0.01 & 12303–40738   & –1.24 $\pm$ 0.03 \\
		       	    & 				   & 40738–81283   & –2.16 $\pm$ 0.05 \\	
	\noalign{\smallskip}
	\hline
\end{tabular}
\end{table}	

The values of the slopes for radial profiles of surface brightness of the coma and tail of comet C/2014 A4 are presented in Table \ref{tab:2}. In the case of a steady-state and free expansion of long-living grains, the $ n $ value in the dependence $I \propto \rho ^n$, which describes the brightness variations with cometocentric distance $\rho$, should be --1. The table shows that the intensity decreases with distance from the nucleus is not uniform. The slopes change with cometocentric distance, and this change is different for the g-sdss and r-sdss filters. Since the slopes differ significantly for the coma and tail, and they also change with increasing distance from the nucleus, we can assume that there is a non-isotropic dust emission from the nucleus, which forms some structures in the innermost coma and tail.

\begin{figure}[t]
	\centering
	\includegraphics[width=0.8\linewidth]{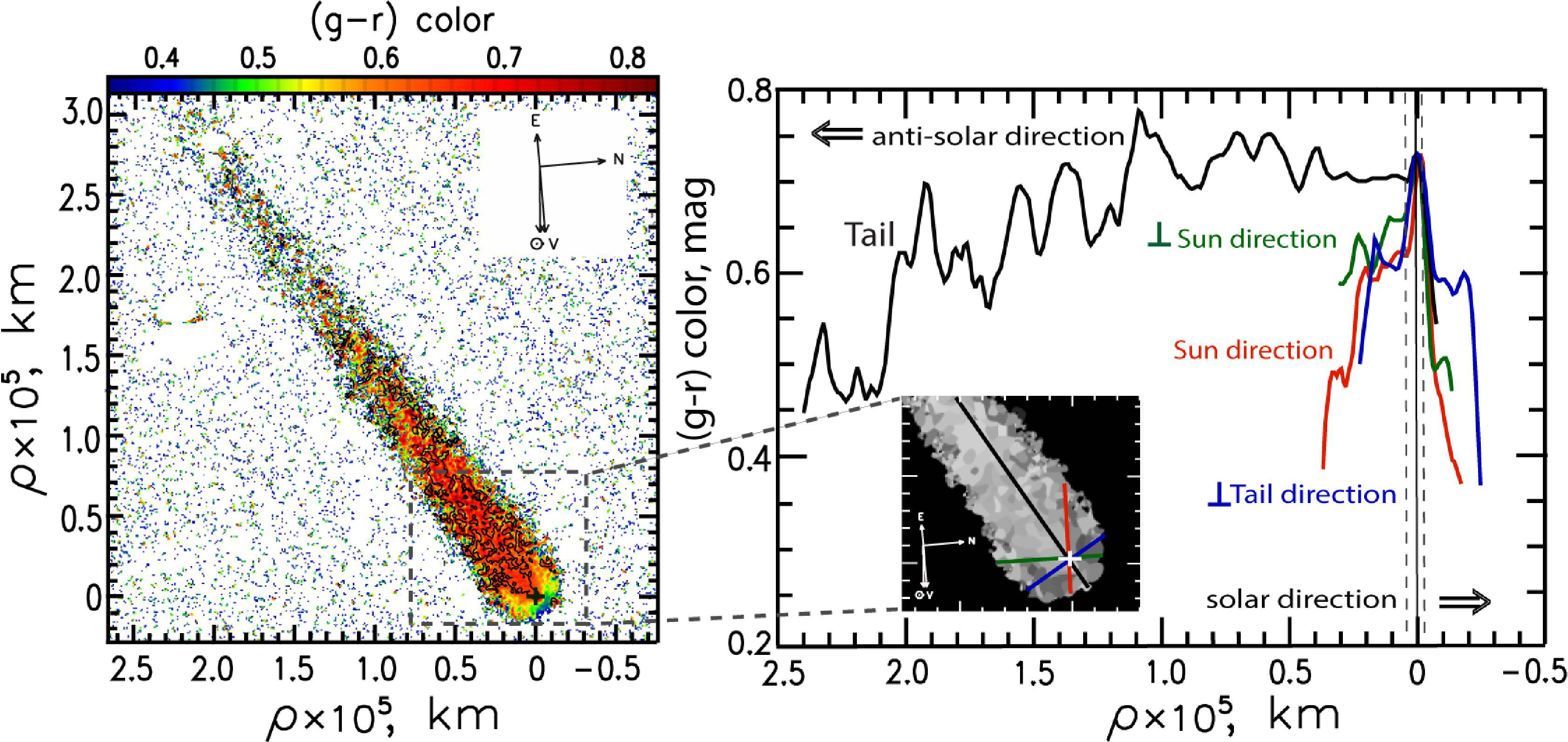}
	\caption{The (g–r) color map of comet C/2014 A4 (SONEAR) derived from the November 5.85, 2015 images in the g-sdss and r-sdss filters (left). The map are colored according to the color indexes in magnitudes as indicated in the bars on the top of the left image. The scans across the color map are displayed on the right: along the solar and anti-solar directions; perpendicularly to the solar-anti-solar direction, and along and perpendicularly to the tail. Positive distance is in the antisolar direction, and negative distance is in the solar direction. The insert shows a larger scale near-nucleus area with the directions of the scans. The vertical dashed lines indicate the spatial extent associated with the seeing. North, East, sunward, and velocity vector directions are indicated.}
	\label{Fig:03}
\end{figure}

\subsection{Color and dust productivity}

Using photometric observations, we  determined the magnitude, dust color, and dust productivity (in the sense of $Af\rho$) for both observation dates. The cometary magnitude can be computed from the following expression:
\begin{equation}
m_c = -2.5 \log_{10} \left[ \frac{I_c(\lambda)}{I_s(\lambda)} \right] + m_s - 2.5 \log_{10} p(\lambda)\Delta M  ,\,
\end{equation}
where $m_c$ is the comet integrated magnitude calculated for the aperture of radius $\rho$, $I_c$ and $I_s$ are the measured fluxes of the comet and the standard star in counts, respectively, $m_s$ is the star magnitude, $p(\lambda)$ is the sky transparency that depends on the wavelength, $\Delta M$ is the difference between the airmasses of the comet and the star. 

To create a color map of the dust coma and tail (Fig. \ref{Fig:03}, left), we converted each pixel for summed images g-sdss and r-sdss into the apparent magnitude and created the final (g--r) color map by subtracting the two sets of images from each other. An average error in the magnitude measurements is 0.04$^m$. Analyzing the color map (Fig. \ref{Fig:03}), one can see that the dust color  (g-r) of comet C/2014 A4 is mainly red, especially in the tail direction. The (g-r) color varies from 0.75$ \pm $0.05$^m$ in the near-nucleus area to 0.45$ \pm $0.06$^m$ along the tail. But in the anti-tail direction, the comet showed another color distribution. In this region, the color of the dust is mostly neutral ($\sim$0.45$^m$) in comparison with the Sun color ((g--r) for the Sun is 0.44$^m$, \href{http://www.sdss.org/dr12/algorithms/ugrizvegasun/}{http://www.sdss.org/dr12/algorithms/ugrizvegasun/}). Digital processing of the images showed a fan-like structure (see Fig. \ref{Fig:01}) in this area.

\begin{figure}[b!]
	\centering
	\includegraphics[width=0.7\linewidth]{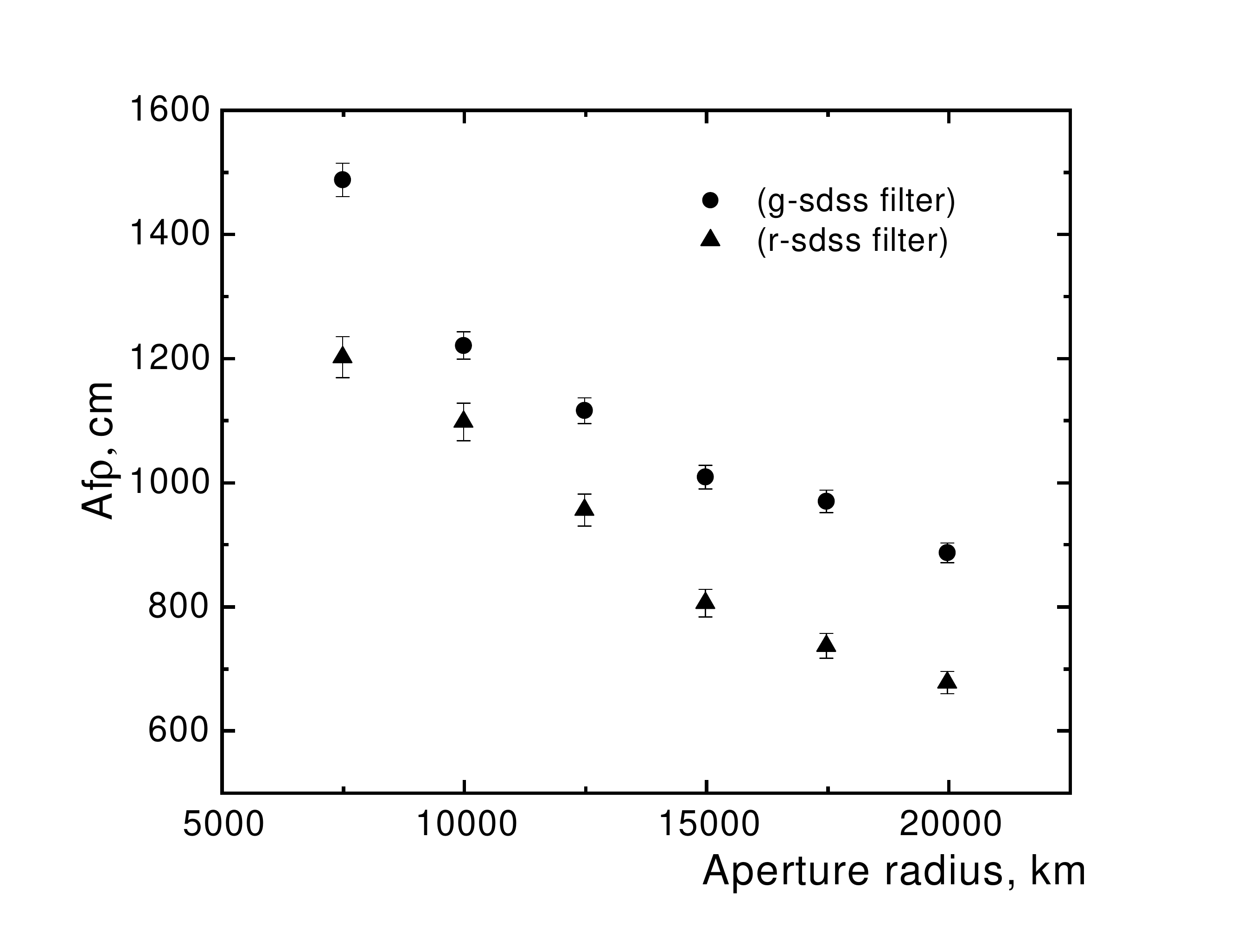}
	\caption{Parameter $Af\rho$ versus aperture radius measured in comet C/2014 A4 (SONEAR) trough the g-sdss and r-sdss filters with BTA telescope. }
	\label{Fig:04}
\end{figure}

We also estimated the parameter $Af\rho$, which was introduced by \citetads{1984AJ.....89..579A}. This parameter characterizes the dust production rate of a comet and is determined by the ratio of the effective scattering cross-section for all the grains entering the field of view of the detector to the projection of this field of view onto the celestial sphere. To calculate values of $Af\rho$, we used comet magnitudes obtained with different aperture radii. For the aperture radius $\sim$20~000~km, the comet magnitude 15.45$\pm$0.05$^m$  was derived for g-sdss filter, and 16.12$\pm$0.03$^m$ for r-sdss filter. Using observations at 6-m telescope in the r-sdss band, we calculated $Af\rho$~=~(680$\pm$18)~cm for an aperture size of 20~000~km.  The dust productions in broadband V and R filters (MASTER-II telescope) were estimated as $Af\rho$ = 659$\pm$39~cm and 645$\pm$36~cm, respectively, for the same aperture radius that close to results obtained at 6-m telescope. 

This value of dust productivity is similar to the value estimated in \href{http://www.astrosurf.com/cometas-obs/C2014A4/afrho.htm}{http://www.astrosurf.com/cometas-obs/C2014A4/afrho.htm} for comet C/2014 A4. Also our results are close to the data obtained by Mauro Facchini (Cavezzo Observatory and Celado Observatory) and Rolando Ligustri (CAST)  see \href{http://cara.uai.it}{http://cara.uai.it}. Their data for comet C/2014 A4, observed on November 7 2015, indicate that the $ Af\rho $ parameter varied from 738 to 783~cm with errors from 27 to 29~cm for R filter. Fig. \ref{Fig:04} shows the variations of $Af\rho$ with the aperture radius in g-sdss and r-sdss filters. In our opinion, this difference results from the difference in the bandwidth and the central wavelength of used filters, i.e. $Af\rho$ reflects contribution of different dust particles. This conclusion is confirmed by our model calculations (see Section 4).
 
We did not calculate the dust mass production rate in this comet because, as showed in \citetads{2018Icar..313....1I}, the result of such calculations strongly depends on the physical characteristics of the dust.  Variations of the mentioned characteristics can lead to dramatic changes in the evaluation of the dust mass production. Although we have obtained some parameters of the dust from modelling (for example, size and composition), we have no information about the velocity of particles. Also we do not know about main gas driver, which controlled of dust outflow, for this distant comet.

\subsection{Spectra}

\begin{figure*}[t!]
	\centering
	\includegraphics[width=0.8\linewidth]{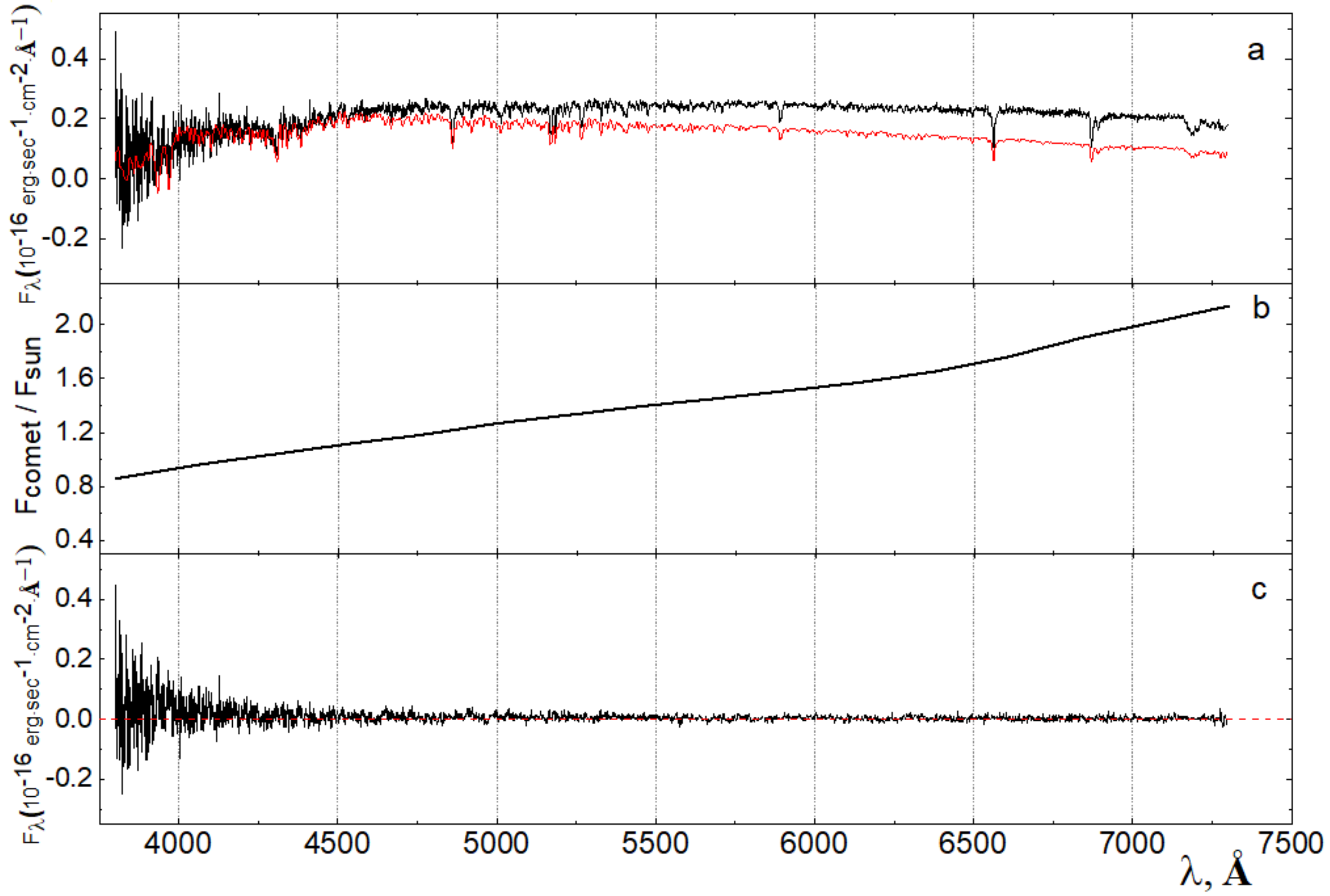}
	\caption{The results of processing the spectrum of comet C/2014 A4 (SONEAR): (a) the energy distribution in the observed cometary spectrum (black line) and the normalized spectrum of the Sun (red line); (b) the normalized spectral dependence of the dust reflectivity; (c) the emission component in the cometary spectrum. }
	\label{Fig:05}
\end{figure*}

Cometary spectra consist of continuum caused by the scattering of sunlight by dust particles and emissions due to reemitting of the solar radiation by molecules in the cometary coma. We used a high-resolution solar spectrum \citepads{1984SoPh...90..205N} to separate these components. The solar spectrum was transformed to the resolution of our observations by convolving with the appropriate instrumental profile and normalized to the flux from the comet. Fig. \ref{Fig:05} shows step by step processing of the comet C/2014 A4 spectrum. The convolved solar spectrum is compared with the cometary spectrum in Fig. \ref{Fig:05}a.  We used multiplication of the solar spectrum by a polynomial (Fig. \ref{Fig:05}b). Subtracting the calculated continuum from the observed spectrum, we obtained a pure emission spectrum (Fig. \ref{Fig:05}c) which does not show any emission features.

\begin{table}[b!]
	\centering
	\caption[]{Upper limits for the main emissions in comet C/2014 A4 (SONEAR).}
	\label{tab:3} 
	\begin{tabular}{lcccc}
		\hline
		\noalign{\smallskip}
		Mol. & $\lambda$/$\Delta \lambda$ & $\sigma \times$10$^{-17}$ & $F\times$10$^{-17}$ & $Q\times$10$^{24}$  \\
		& $\AA$&   erg s$^{-1}$cm$^{-2} \AA^{-1}$   & erg s$^{-1}$cm$^{-2}$    &    mol s$^{-1}$    \\
		\noalign{\smallskip}
		\hline
		\noalign{\smallskip}
		CN     &	3870/62	& 10.75	& < 55.91 & < 9.05 \\
		C$_3$  &	4062/62	& 5.15	& < 26.78 & < 0.15 \\
		CO$^+$ &	4266/64	& 2.36	& < 12.29 & 	\\
		C$_2$  &	5141/118 & 1.24	& < 6.47	 & < 0.98 \\
		\noalign{\smallskip}
		\hline
	\end{tabular} 
	
\end{table}

We determined upper limits to the emission fluxes of CN, C$_3$, C$_2$, CO$^+$, and upper limits to their production rates. For this, we adopted a Gaussian function having a FWHM equal to the spectral resolution as an equivalent of the minimal measurable signal. The amplitude of the Gaussian was equal to the RMS noise level calculated within wavelength regions associated with the bandpass of the narrowband cometary filters \citepads{2000Icar..147..180F}. The Haser model \citepads{1957BSRSL..43..740H} was used to derive the upper limits to the production rates of the neutrals. The model parameters were taken from \citetads{2011Icar..213..280L}. The upper limits to the fluxes ($ F $) and gas production rates ($ Q $) are listed in Table \ref{tab:3}. These values have been computed with aperture of 1.56$^{\prime\prime}$.

We also investigated variations of the reflectivity $S(\lambda)$ along the dispersion expressed as the comet spectrum $F_{com}(\lambda)$ divided by the scaled solar spectrum, $F_{sun}(\lambda)$: $S(\lambda)=F_{com}(\lambda)/F_{sun}(\lambda)$. We used the polynomial from Fig. \ref{Fig:05}b to derive the dust reflectivity. The result shows a growth of dust reflectivity with increasing wavelength and can be presented quantitatively as the normalized gradient of reflectivity using the following expression \citepads{1984AJ.....89..579A}
\begin{equation}
S'(\lambda_1,\lambda_2)  = \frac{2}{\lambda_2 - \lambda_1} \frac{S(\lambda_2) - S(\lambda_1)}{S(\lambda_2) + S(\lambda_1)}  ,\,
\end{equation}
where $S(\lambda_2)$ and $S(\lambda_1)$ correspond to the measurements at the wavelengths $\lambda_1$ and $\lambda_2$ (in $\AA$) with $\lambda_2$>$\lambda_1$. $S^\prime (\lambda_1,\lambda_2)$ is expressed in percent per 1000~$\AA$. With the adopted first degree of the polynomial fitting, the derived reddening is equal to 21.6$\pm$0.2\% per 1000~$\AA$ and is valid for the whole examined wavelength region that is 4650~--~6200~$\AA$.

For comparison, \citetads{1992Icar...98..163S} obtained the mean value of about 22\% per 1000~\AA~(with the minimum and maximum values of 15\% per 1000~\AA~ and 37\% per 1000~\AA, respectively) within the 4400 -- 5600~\AA~wavelength region for 18 ecliptic comets. For dynamically new distant comets, \citetads{2018A&A...611A..32K} obtained rather different results: the normalized gradient of reflectivity was about 12\% per 1000~\AA~and about 7\% per 1000~\AA~ for BV and VR spectral regions, respectively. Thus, the color of comet C/2014 A4 is comparable with the color of ecliptic comets, but is significantly redder than the color of the distant comets studied by \citetads{2018A&A...611A..32K}. To find out a reason of this difference, a systematic study of the color of distant comets is required, especially because the color of comets may change significantly even on a daily basis \citepads[see Fig. 2  in][]{2004come.book..577K,2017MNRAS.469.2695I,2019MNRAS.tmp..658L}.

\begin{figure}[t!]
	\centering
	\includegraphics[width=0.9\linewidth]{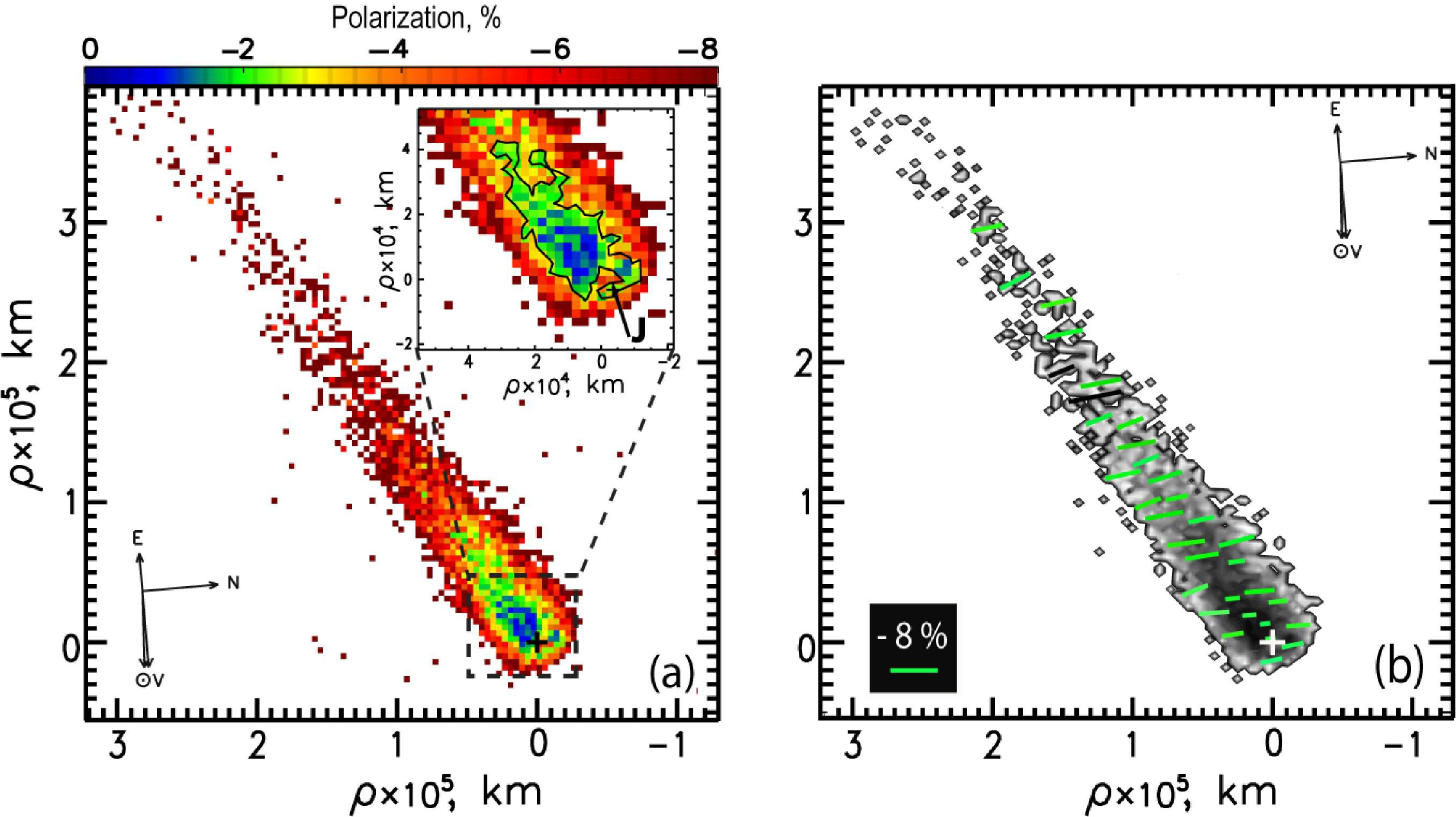}
	\caption{Distribution of linear polarization degree (a) and polarization vectors (b) over the comet C/2014 A4 (SONEAR) coma derived  through the Rc filter. On the left image indicate the polarization with its associated scale bar in percentage on the top of the image. The location of the optocenter is marked with a white cross. The orientation of the vectors indicates the direction of the local polarization plane, and the length of the vectors indicates the degree of polarization. The projected directions toward the Sun, North, East, and comet velocity vector (V) are indicated.  }
	\label{Fig:06}
\end{figure}

\subsection{Polarimetry}

The map of polarization degree for comet C/2014 A4, obtained in the broad-band R$_c$ ($\lambda$6420/790~$\AA$) filter at the 6-m telescope, is presented in Fig. \ref{Fig:06}a. A fan-shaped region near the  sunward direction,  also observed in photometric images (Fig. \ref{Fig:01}), has a higher polarization then the surrounding coma. It seems that there is also a structure with a lower polarization extending from the near-nucleus region in the direction of the tail.

\begin{figure}[t!]
	\centering
	\includegraphics[width=1.0\linewidth]{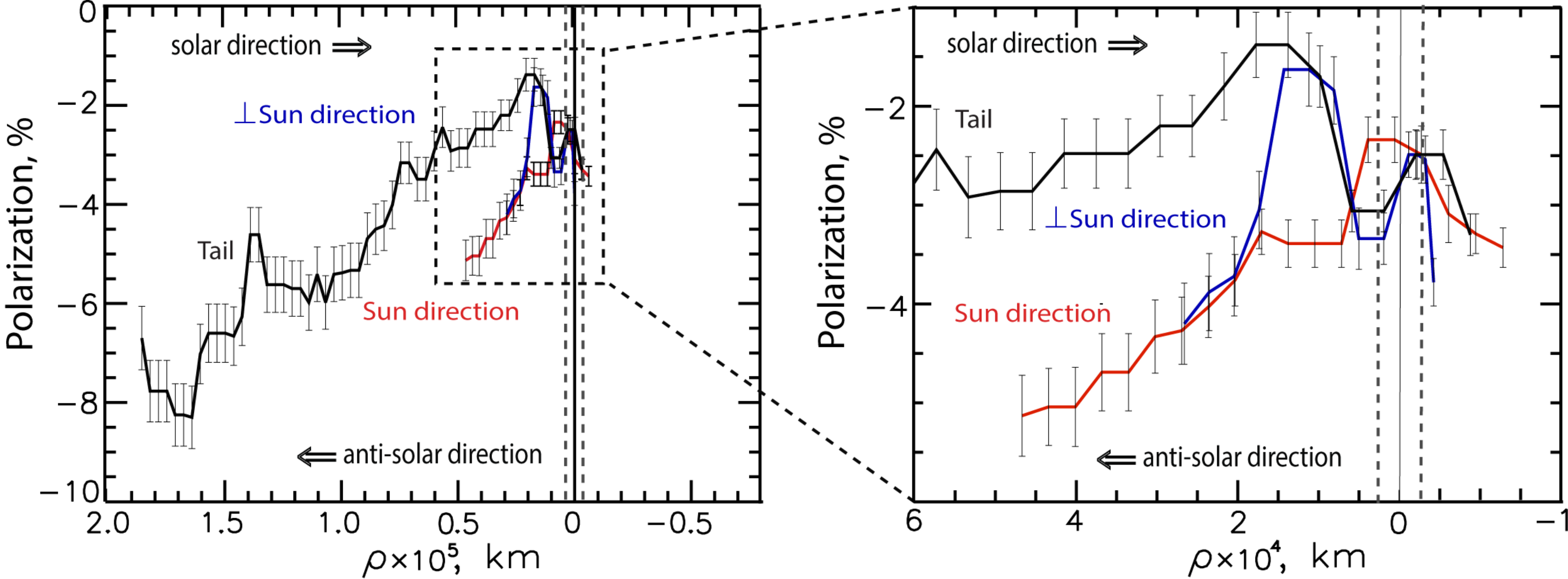}
	\caption{The radial profiles of polarization across the coma of comet C/2014 A4 (SONEAR). On the left side, there are the full polarization profiles, and on the right side, the profiles for the near-nucleus area defined by the dashed rectangle are shown. Vertical doted lines indicate the seeing size. }
	\label{Fig:07}
\end{figure}

Figure \ref{Fig:07} shows the observed radial profiles of polarization across the coma in the solar and perpendicularly to the solar-anti-solar directions and in the tail direction. Measurements of the polarization were performed with a 3$\times$3~px size aperture along the selected directions in the coma. The maximum degree of polarization along the tail and perpendicular to the sunward direction was shifted relative to the photometric center toward the tail.

As one can see in Fig. \ref{Fig:07}, there are significant differences in the radial profiles of polarization along the tail, in the solar direction, and perpendicular to the solar direction. In the near-nucleus region, $\pm$10~000~km from the optocenter, all profiles display polarization variations  within the range from --2.5\% to --3.5\%. At the distances 10~000~---~20~000 km, the polarization dropped (by absolute value) sharply, from approximately --3.7\% to --1.5\%. Then, the degree of polarization gradually, although with different gradients, increases, reaching about --5\% at cometocentric distance 60~000~km in the solar direction and --8\% at about 170~000~km in the tail. Note that the changes in polarization are not symmetrical relatively to the photometric centre.

The absolute values of the linear polarization for comet C/2014 A4 are significantly higher than those ever observed in the comets at heliocentric distances less than 2 au, where typical values do not exceed 2\% \citepads{2015psps.book..379K}. Our previous observations of distant comets C/2010 S1 (LINEAR), C/2010 R1 (LINEAR), C/2011 KP36 (SPASEWATCH), C/2012 J1 (CATALINA), C/2013 V4 (CATALINA)  \citepads{2015P&SS..118..199I,2017EPSC...11..101I} also showed the  negative polarization degree whose absolute values  changed with the distance from the nucleus from $ \sim $~--1.5 to $\sim$~--8\%.

The distribution of the polarization vector (Fig. \ref{Fig:06}b) does not indicate any alignment of the dust particles. On average, the polarization vector have about $ PA =80 \pm 4^\circ$, i.e., practically parallel to the scattering plane ($\phi$=85$^\circ$). This orientation of the polarization plane is typical in the case of negative polarization, i.e., for the light scattered by dust at the phase angles < 20$^\circ$. 

\begin{table*}[b!]
	\centering
	\caption[]{Complex refractive index of the materials considered in this study.}
	\label{tab:4} 
	\begin{tabular}{llll}
		\hline
		\noalign{\smallskip}
		Material & $ m $ & $ m $ & Reference  \\
		& g-sdss filter & r-sdss filter &     \\
		\noalign{\smallskip}
		\hline
		\noalign{\smallskip}
		Halley-like dust &	1.88 + \textit{i}0.47 & 1.98 + \textit{i}0.48 &	\citetads{2003AA...407L...5K} \\
		Amorphous silicate (forsterite) & 	1.689 + \textit{i}0.0031 & 1.677 + \textit{i}0.0044 & \citetads{1996ApJS..105..401S} \\
		Water ice  & 	1.3157 + \textit{i}1.54$\times$10$^{-9}$ & 1.308 + \textit{i}1.43$\times$10$^{-8}$ & \citetads{1984ApOpt..23.1206W}  \\
		Carbon dioxide ice 			&	1.42 + \textit{i}7.04$\times$10$^{-7}$ & 1.41 + \textit{i}1.05$\times$10$^{-6}$ &  \citetads{1986ApOpt..25.2650W} \\
		Iron-rich pyroxene 			&	1.702 + \textit{i}0.0596 & 1.675 + \textit{i}0.0212 & \citetads{1995AA...300..503D} \\
		Olivine						& 1.777 + \textit{i}0.126 &	1.743 + \textit{i}0.066 & \citetads{1995AA...300..503D} \\
		Amorphous carbon 			&	1.95 + \textit{i}0.786 & 2.14 + \textit{i}0.805 & \citetads{1991ApJ...377..526R} \\
		Pyrrhotite (FeS) 			&	1.45 + \textit{i}1.53 &	1.70 + \textit{i}1.86 & \citetads{1977Icar...30..413E} \\
		Fe-rich olivine (Fayalite) 	&	1.85 + \textit{i}0.0015 & 1.85 + \textit{i}0.00077 & \citetads{2001AA...378..228F} \\
		Cosmic organic refractory 	&	1.9146 + \textit{i}0.3173 &	1.98 + \textit{i}0.2677 & \citetads{1997AA...323..566L} \\
		Tholin ice 					&	1.56 + \textit{i}0.0025 & 1.540 + \textit{i}0.0012 & \citetads{1993Icar..103..290K}
		\\
		Titan’s tholin 				&	1.58 + \textit{i}0.0058 & 1.557 + \textit{i}0.009 & \citetads{2002Icar..156..515R} \\
		\noalign{\smallskip}
		\hline
	\end{tabular} 
	
\end{table*}

\section{Modeling of variations of polarization and color over the coma}

To interpret the observational data, we used the rough spheroids model described in \citetads{2015P&SS..116...30K}. There we found that the model can provide a good fit to the photopolarimetric observational data for comets using realistic characteristics of the dust particles. The rough spheroids model represents the dust as a polydisperse ensemble of spheroids with a range of aspect ratios. Roughness is presented by a normal distribution of the surface slopes, and is defined by the standard deviation of the distribution, which is zero for smooth surface and greater than zero for rough surface. A great advantage of the model is that there is a library of pre-calculated kernels for computations of the light scattering characteristics of rough spheroids \citepads{2006JGRD..11111208D}, using which one can quickly obtain brightness and polarization for a variety of dust compositions, size distributions, and spheroid shapes. 

\subsection {Properties of the particles producing the observed negative polarization and red color}

Based on the best fit for the cometary data obtained in \citetads{2015P&SS..116...30K}, we selected a log-normal size distribution of particles with the effective variance $v_{eff}$~=~0.1 and the range of aspect ratios of spheroids from 0.3 to 3, thus, covering both prolate and oblate spheroids. We described the spheroid roughness by the standard deviation $\sigma$~=~0.2 that was the maximum roughness available in the pre-calculated kernels. The refractive indexes $ m $ for the materials considered in this study are listed in Table \ref{tab:4}. There Halley-like dust represents a composition that was found typical for comet 1P/Halley dust particles based on Giotto mass-spectrometer data \citepads{1988Natur.332..691J}. Note that now we have new data on the composition of cometary dust resulted from the Rosetta studies of comet 67P/Churyumov--Gerasimenko \citepads{2017MNRAS.469S.712B}, these data do not affect the refractive index of the dust substantially, and our computations show that the light scattering characteristics of the 1P/Halley and 67P/Churyumov--Gerasimenko dust look almost identical \citep{2018Kolokolova}.

\begin{figure*}[t!]
	\centering
	\includegraphics[width=1.0\linewidth]{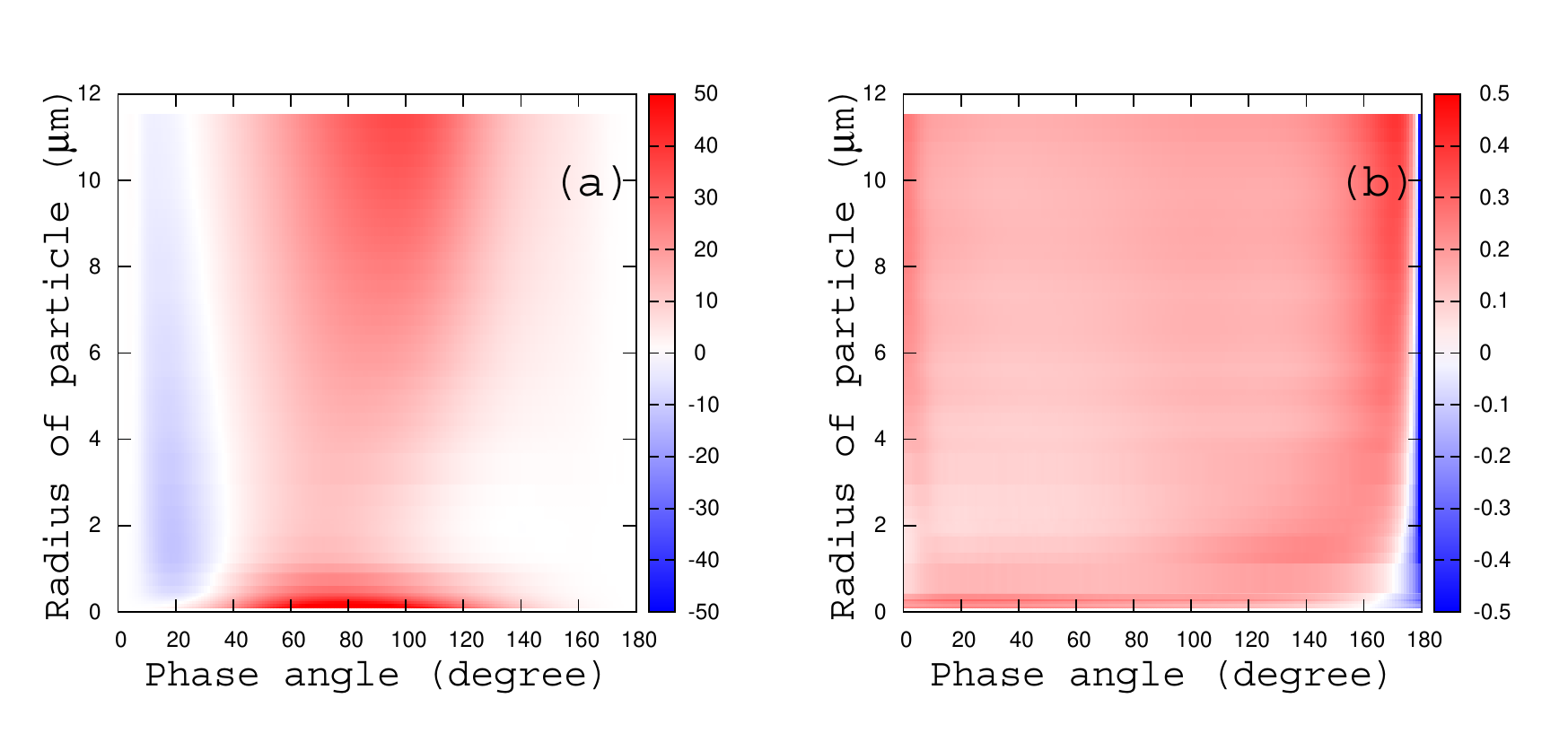}
	\caption{Polarization degree is in \% (left) and (g--r) color (right) images computed for a mixture of porous Halley-like particles (porosity 85\%) and solid silicate particles (forsterite); the ratio of the constituents in the mixture is 9:1. The figure demonstrates applicability of the rough spheroid model to modeling polarization and color for regular (not distant)  comets. Polarization is given in percents, and color is in magnitudes but calculated for the white-light illumination, i.e., they represent the intrinsic color of the dust, or, from astronomical point of view, the dust color excess.}
	\label{Fig:08}
\end{figure*}

Based on the success of modeling cometary photopolarimetric data in \citetads{2010EP&S...62...17K} and \citetads{2015P&SS..116...30K}, we started with the same model as in those papers, i.e., considered cometary dust as a mixture of porous Halley-dust and solid silicate particles. Although this model allows us to correctly reproduce the regular polarization phase curve and color (Fig. \ref{Fig:08}) and, for some size distributions, to reach negative polarization as low as $-$11\% at phase angles $\sim$15--18$^\circ$, with this composition of cometary dust we could not reach polarization smaller than $-$0.3\% at phase angle $\sim$4$^\circ$ for the size distributions of any  effective radius of particles. Numerous simulations for this mixture showed that changing parameters of the size distributions or considering Halley dust of different porosity or containing different silicates from Table \ref{tab:4} did not improve the results. Adding carbon or organics only worsens the situation as those absorptive materials led to vanishing negative polarization.

\begin{figure*}
	\centering
	\includegraphics[width=1.0\linewidth]{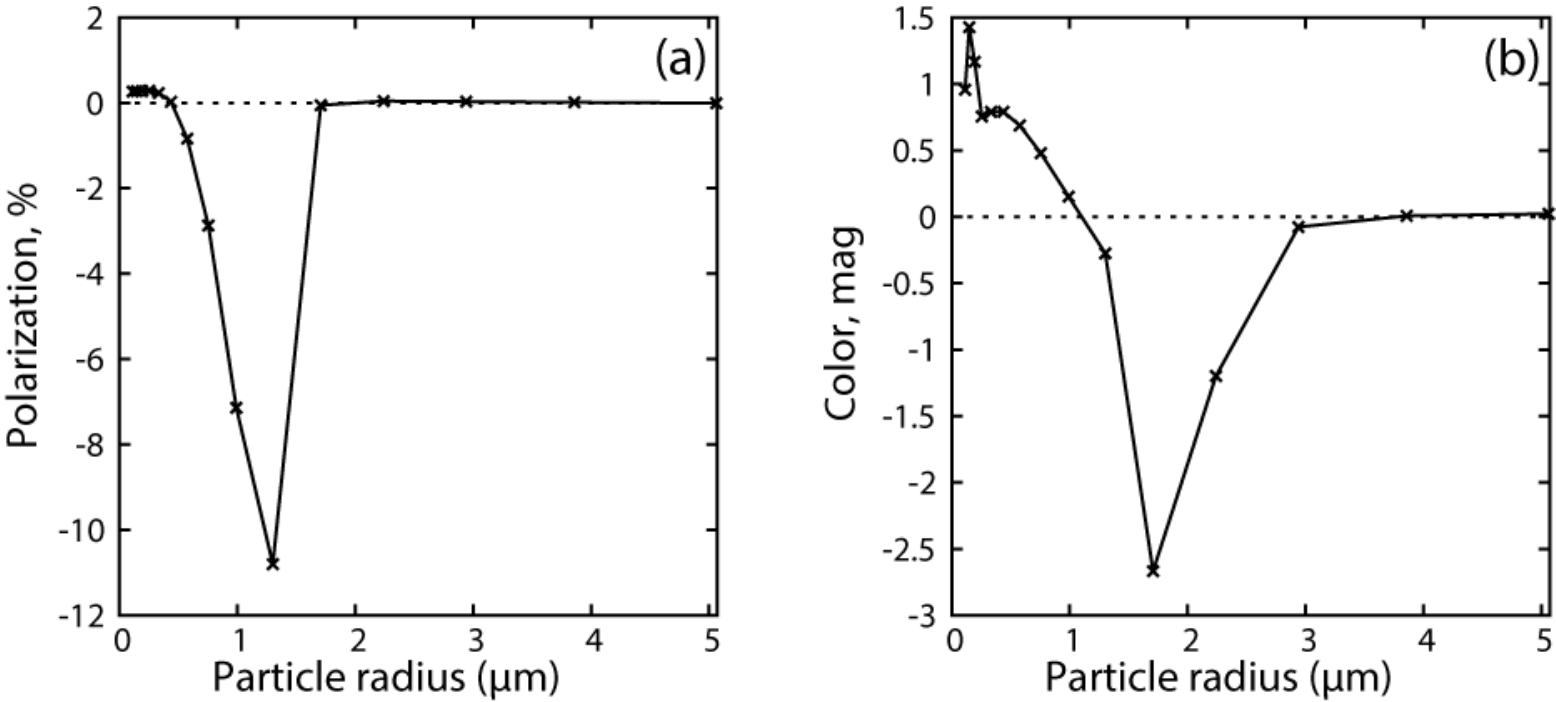}
	\caption{ Polarization (a) and (g--r) color (b) for ice of porosity 55\% at phase angle 4$^\circ$. Polarization was computed for r-sdss filter. Here we use the same definition of color as in Fig. \ref{Fig:08}, i.e., the presented values are intrinsic colors of the considered particles and are analogous to  their color excess. }
	\label{Fig:09}
\end{figure*}

We accomplished computer simulations of polarization and color for the materials listed in Table \ref{tab:4}. Our results showed that none of the absorbing materials (imaginary part of the refractive index >0.1) produced a negative polarization. Thus, to reproduce the observed negative polarization,  we needed some transparent material, i.e. a material with a low imaginary part of the refractive index. Our modelling for the silicates presented in Table \ref{tab:3} did not show negative polarization more negative than --0.7\% at phase angle 4$^\circ$. Besides, it was hard to imagine a comet with the dust made of pure silicates. The most applicable candidate for the comets observed at large heliocentric distances is water ice, and our computations showed that particles made of pure water ice produce polarization about --3\% at 4$^\circ$. The lowest polarization at 4$^\circ$ was achieved for porous ice with porosity 55\% (Fig. \ref{Fig:09}a) whose refractive index was calculated for a mixture of ice and voids using the Maxwell Garnett mixing rule. However, this and any other porous or solid ice showed a very blue color (Fig. \ref{Fig:09}b) at small phase angles. We also tried carbon dioxide ice and received similar results. However, since carbon dioxide was not detected in comet C/2014 A4 by NEOWISE observations (Bauer, private communication), we continued our modeling using water ice.
 \begin{figure*}[b!]
 	\centering
 	\includegraphics[width=1.0\linewidth]{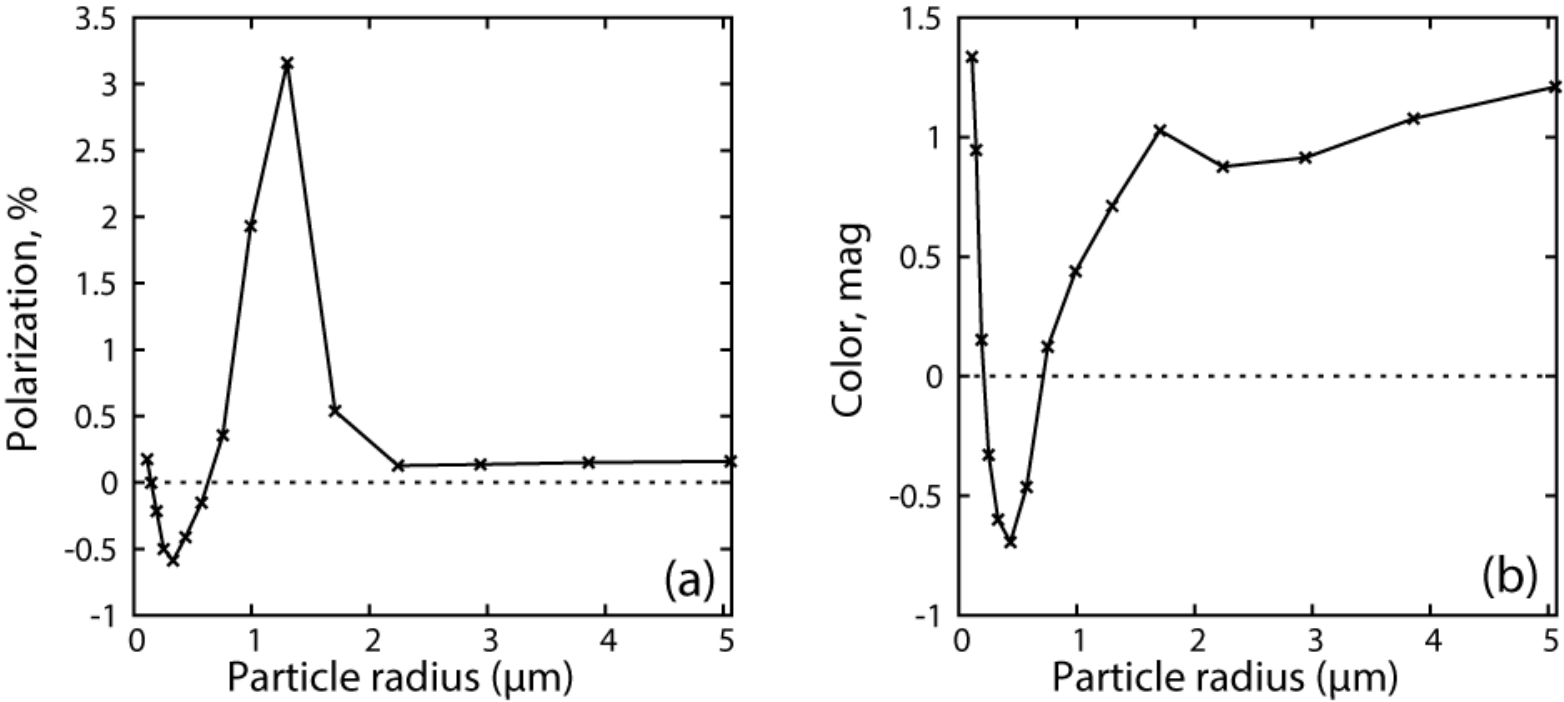}
 	\caption{The same as Fig. \ref{Fig:09}, but for Titan tholins.}
 	\label{Fig:10}
 \end{figure*}
 To reproduce the observed red color, we needed a very red material, whose redness could overpower the blue color of ice. However, this material should not produce a significant positive polarization that might cancel the ice negative polarization, i.e., it should have a low imaginary part of the refractive index. After numerous attempts, we found that these characteristics were typical for tholins, and our computations showed (Fig. \ref{Fig:10}) that Titan’s tholin had the necessary photopolarimetric properties for particles larger than 1~$\mu$m. holins as a component of cometary material has been already considered in a number of publications starting from pioneering papers by \citetads[][and other papers by those authors]{1979Natur.281..708S} and up to the papers related to interpretation of the Rosetta data (e.g., \citeads{2015Icar..256..117S,2016Icar..267..154P,2019MNRAS.484.2198F}). Besides,   \citetads{2015Icar..250...53B} showed that IOM (insoluble organic matter) found in carbonaceous meteorites and also considered as the organic component in the Rosetta dust  \citepads{2017MNRAS.469S.712B} could originate from tholin-like organics. Although tholin in comets may not be identical to Titan’s tholin, we have found that spectral properties of Titan’s tholin are similar to those of other (e.g., Triton, see \citeads{1994Icar..108..137M}) tholins, and Titan’s tholin is the only type of tholins with reliably measured optical constants; it was used in many papers to interpret cometary, including Rosetta (see references above), data.  We also performed computations for tholin ice, and the results were very similar to the Titan’s tholin that is not surprising from comparison their refractive indexes in Table \ref{tab:4}.

From Figs. \ref{Fig:09}~--~\ref{Fig:10}, one can see that combining icy particles of size between 1.5 and 2~$\mu$m with tholin particles of size >2~$\mu$m, we can reach very negative values of polarization and red color. This combination may be unrealistic for the comets observed at the heliocentric distances about 1~au, however, it sounds quite reasonable for the distances around 4~au where comet C/2014 A4 was observed.

\subsection{Simulating the observed polarization and color trends in the coma}

For a fair interpretation of our data, the found characteristics of the dust particles in the tail should be consistent with the values of polarization and color observed near the nucleus. This means that the polarization should be around $-$2\% and color should be more red than in the tail, but the composition of the dust particles near the nucleus should be similar to the composition of the particles in the tail and the size of particles near the nucleus should be similar or larger since on the way to the tail particles can only evaporate, fragment or be sorted by radiation pressure the way that small particles become more abundant in the tail. 

An extensive analysis of all the data we have obtained during our simulations with the goal to find a behavior of the dust that can reproduce the values of polarization and the observed trends with realistic types of particles resulted in the following characteristics of the dust in comet C/2014 A4:
\begin{enumerate}
\item	Polarization in the tail, $ - $8\% can be achieved by combination of tholin particles of radius >3~$\mu$m and porous icy particles of radius >0.75~$\mu$m (porosity 30~--~50\%) with the abundance of icy particles >98\% in the mixture.
\item	Polarization in the non-tail directions, far from the nucleus, equal to --5\%, needs tholin particles of radius >3~$\mu$m and porous icy particles of radius >1.0~$\mu$m (porosity $\sim$30\%). Abundance of icy particles should be 85~--~90\% in the mixture.
\item	Polarization near the nucleus, --2.5\%, can be reproduced by tholin particles of radius >3~$\mu$m, mixed with porous icy particles of radius <1.3~$\mu$m (porosity 30~--~50\%). The larger are the icy particles the smaller abundance of them is required. For example, icy particles of radius $\sim$1~$\mu$m require 55\% of them in the mixture, and even less for larger particles.
\end{enumerate}

\begin{table*}[h!]
	\centering
	\caption[]{The best-fit characteristic of the dust particles from our modeling}
	\label{tab:5} 
	\begin{tabular}{lllllll}
		\hline
		\noalign{\smallskip}
		Region   & Icy particles  & Icy particles & Tholin particles & Tholin particles & Polarization     & Color  \\
		of comet & radius           &  abundance    & radius    		& abundance         & in r-sdss filter &  (g--r),  \\
		& ($\mu$m)	      &   (\%)        & ($\mu$m)    	& (\%)              & (\%)			   & (mag)  \\
		\noalign{\smallskip}
		\hline
		\noalign{\smallskip}
		\multirow{2}{*}{Near the nucleus} & 1.3  & 40~---~50 & 3~---~5 & 60~---~50 & --~2.7~---~--~2.0 & 0.85~---~0.91 \\
		& 1.0  & 65~---~70 & 3~---~5 & 35~---~30 & --~2.1~---~--~2.0 & 0.53~---~0.47 \\
		\noalign{\smallskip}
		\hline
		\noalign{\smallskip}
		Non-tail 
		directions & 1.0  &	85~---~90 & 3~---~5 &	15~---~10 & --~5.3~---~--~6.3 & 0.66~---~0.56 \\
		\noalign{\smallskip}
		\hline
		\noalign{\smallskip}
		In the tail& 0.75 &	98~---~99.5& 3~---~5 & 2~---~0.5 &	--~8.5~---~--~8.25 &	0.48~---~0.52 \\
		\noalign{\smallskip}
		\hline
	\end{tabular} 	
\end{table*}

In Table \ref{tab:5}, the solar color 0.44$ ^m $ was added to the calculated values of color to make them comparable with the observed that ranged from $\sim$ 0.7 near the nucleus  to $\sim$ 0.4 in the tail; abundance means percent of the icy (or tholin) particles in the dust, i.e. describes their contribution to the number density of particles. One can see that with our modeling we managed to reproduce the observed values of polarization and color, as well as their changes with the distance from the nucleus.  The key result here is the combination of a very negative polarization with a very red color that was the main observational finding for this comet. We also successfully reproduced their significant change with the distance from the nucleus. Since the results were obtained with a rather simple model of rough spheroids, some small deviations from the observed values are not surprising.   Note that adding a small amount of silicates (e.g., as a material underlying the tholin layer) can slightly  decrease the color (making it even closer to the observed values) without affecting the values of polarization.

It may look doubtful that there was a smaller number of icy particles near the nucleus (e.g., 55\%) than in the tail (>98\%) if we assume that the particles can only change due to their sublimation. However, if we assume that the particles fragment on their way out of nucleus and in the case of the tail direction also sorted by radiation pressure, then we should expect that the number density of small icy particles increases with the distance from the nucleus. This can explain not only the smaller size of icy particles far from the nucleus, but also the larger abundance of icy particles there, especially in the tail. Notice that the size of the tholin particles is not changing, i.e. their abundance, relatively to the icy particles, should decrease with the distance.

The conclusion about the fragmentation of the particles in the coma is confirmed by our photometric data. For example, the decrease in $Af\rho$ with the distance from the nucleus (Fig. \ref{Fig:04}) most naturally can be explained the fragmentation of the particles as fragmentation causes a decrease in particle scattering cross-section. We see indication of fragmentation also in the radial profiles (Fig. \ref{Fig:02}). The profile curves are mostly steeper than 1/$\rho$ that confirms decrease in particle size (see Section 3.2). Finally, we see a noticeable difference in the radial profiles for r-sdss and g-sdss filters. Since the light scattering characteristics of particles change quickly when the particle size is comparable to the wavelength, this confirms the predominance of submicron and micron-sized particles. We may probably claim that the particles size is close to 1~$\mu$m as the g-sdss filter profiles are usually flatter than r-sdss filter profiles, indicating that the particle fragmentation affects the r-sdss filter, whose wavelength is closer to 1 micron, stronger than the g-sdss filter. Thus, our photometric data are in a good agreement with the results of our modeling.

\section{Conclusions}

We present the results of photometric, spectroscopic, and polarimetric observations of distant comet C/2014 A4 (SONEAR) carried out at the 6-m (SAO RAS) and 0.4-m telescope (Kourovka observatory, Russia) on November 5~--~7, 2015. From the analysis of the spectral observations we did not reveal any emissions at or above the 3$\sigma$ level. We estimated an upper limit of the gas production rate in the comet for the main emissions (CN, C$_3$, C$_2$, and CO$^+$), which for C$_2$ is equal to 0.98$\times$10$^{24}$ mol~s$^{-1}$. The continuum shows a reddening effect with the normalized gradient of reflectivity (21.6$\pm$0.2)\% per 1000~$\AA$ within the 4650~--~6200~$\AA$ wavelength range. We characterized the dust production via $Af\rho$~=~680$\pm$18~cm in the r-sdss filter. 
 
Most of our knowledge about the physical properties of cometary dust were obtained based on the observations of comets close to the Sun (less than 2~au). Also, it was considered that the nature of the dust particles presented in cometary coma does not depend on the heliocentric distance \citepads{1988A&A...206..348D}. However, new investigations show differences between the activity of comets close to the Sun and distant comets \citepads[see,  e.g.,][]{2007MNRAS.381..713M,2008MNRAS.390..265M,2011Icar..211..559I,2015Icar..258...28I,2018JQSRT.205...80D}. Most likely the particles which form the comae are different. Ground-based polarimetric observations of distant comets have not been carried out until recently. The first detailed analysis of distribution of linear polarization in cometary coma for distant comets, with perihelion more than 4~au, were published by \citetads{2015P&SS..118..199I,2015AstBu..70..349I}. Our results for comet C/2014 A4 support these results, which show a deeper branch of negative polarization at small phase angles in comparison with that observed for comets close to the Sun. The only case of a large negative polarization  equal to --6\% was reported by \citetads{2003JQSRT..79..661H} for the circumnucleus halo of comets 81P/Wild 2 and 22P/Kopff. But those comets were observed at phase angles 9.7 and 18$^\circ$ and at those phase anges polarization can reach very negative values even for a rather regular composition of the dust (see Fig. \ref{Fig:08}). \citetads{2014ApJ...780L..32H} found that the average (negative) polarization over the coma was --1.6\% for the comet C/2012 S1 (ISON) at heliocentric distances 3.81~au. Such negative polarization is unusually low in comparison with the observed in other distant comets that may be related to the dynamical uniqueness of comet C/2012 S1 (ISON), which ended up as a sungrazing  comet. In our observations of comet C/2014 A4, polarization map shows spatial variations of the degree of polarization over the coma from about --3\% near the nucleus to almost --8\% in the tail. Analysis of the polarization and color and their change with the distance from the nucleus shows that (1) dust particles in the distant comets are small (<1~$\mu$m) that is not surprising as there is too little gas that can lift large particles; (2) dust contains icy particles and particles composed of (or covered by) tholin or similar organics; (3) icy particles fragment as they move out of the nucleus. The photometric characteristics are consistent with the submicron/micron size of particles and the fragmentation of the dust particles in the coma. 

Thus, our results suggest that the dust in distant comets differ from the dust in regular comets by both size and composition. For the regular comets, usually rather large particles made of a mixture of dark Halley-like material and silicates provide the best fit \citepads{2015psps.book..379K}. For the case of distant comet C/2014 A4, we have found that the dust particles are in the submicron-micron size range and their composition is characterized by a high abundance of ice and tholin-like organics.  Such characteristics of the dust in distant comets are quit realistic as we cannot expect that far from the Sun the amount of sublimating gases can lift as large particles as near the Sun. We can also expect that tholin is transformed close to the Sun, e.g., as described in \citetads{2015Icar..250...53B}. 

Our results show how much more information can be extracted from observational data if we analyze photometric and polarimetric data together, especially combining polarization and color. Also, we enhance our understanding of the evolution of the dust in the coma if the analysis of polarization and color is complemented by studying variations of other photometric characteristics of the dust, specifically, its radial profiles and change in $Af\rho$ with the distance from the nucleus.

\begin{acknowledgements}
The observations at the 6-m BTA telescope were performed with the financial support of the Ministry of Education and Science of the Russian Federation (agreement No. 14.619.21.0004, project ID RFMEFI61914X0004). O.I. thanks the SASPRO Programme No. 1287/03/01 for financial support. The research leading to these results has received funding from the People Programme (Marie Curie Actions) European Union’s Seventh Q12 Framework Programme under REA grant agreement No. 609427. Research has been further co-funded by the Slovak Academy of Sciences grant VEGA 2/0023/18. I.L. thanks the SAIA Programme for financial support. The researches by O.I., I.L. and V.R. are supported, in part, by the project 19BF023-02 of the Taras Shevchenko National University of Kyiv. H.S.D. wants to acknowledge SERB, India for funding travel grant to travel UMD for collaborative work. We thanks C.A.R.A. project, and personally Gianantony Milani, Mauro Facchini (Cavezzo Observatory and Celado Observatory) and Rolando Ligustri (CAST) for provision of their unpublished $Af\rho$ data for comet C/2014 A4 (SONEAR) to comparison in our manuscript.  We thank Ivan Mamoutike and Lennard Poliakov who analyzed the modeling data and compared them with the observations.

\end{acknowledgements}

%
%

\end{document}